\documentclass[prb,preprint]{revtex4} 

\usepackage{amsmath}

\begin{document} 

\title{Adventures in Friedmann cosmology: 
A detailed expansion of the cosmological Friedmann equations}

\author{Robert J. Nemiroff}
\email{nemiroff@mtu.edu}
\author{Bijunath Patla}
\email{brpatla@mtu.edu}
\affiliation{Michigan Technological University, Department of Physics, 
1400 Townsend Drive, Houghton, Michigan 49931}


\begin{abstract}
The general relativistic cosmological Friedmann equations which describe how the scale factor of the universe evolves are expanded explicitly to include energy forms not usually seen. The evolution of the universe as predicted by the Friedmann equations when dominated by a single, isotropic, stable, static, perfect-fluid energy form is discussed for different values of its gravitational pressure to density ratio $w$. These energy forms include phantom energy ($w<-1$), cosmological constant ($w=-1$), domain walls ($w = -2/3$), cosmic strings ($w = -1/3$), normal matter ($w = 0$), radiation and relativistic matter ($w = 1/3$), and a previously little-discussed form of energy called ``ultralight" ($w>1/3$).  A brief history and possible futures of Friedmann universes dominated by a single energy form are discussed. 
\end{abstract}


\maketitle

\section{The Friedmann Equation of Energy, Expanded}

The Friedmann equation of energy for a uniform cosmology evolving under general relativity is typically written in the form\citep{Pea99}
\begin{equation} \label{FriedmannOld}
H^2 = {8 \pi G \over 3} \rho - {k c^2 \over R^2} ,
\end{equation}
where $H$ is the Hubble parameter, $G$ is the gravitational constant, $\rho c^2$ is the energy density, $c$ is the speed of light, $R$ is a scale factor of the universe, and $k$ is a dimensionless constant related to the curvature of the universe. The Hubble parameter $H = {\dot R}/R = {\dot a}/a$, where $a$ is the dimensionless scale factor of the universe such that $a = R/R_0$ and $R_0$ is the scale factor of the universe at some canonical time $t_0$.  An example of $R_0$ is the average distance between galaxies. A derivation of the Friedmann equation of energy and the Friedmann equation of acceleration (discussed in a following section) can be found in numerous sources and is not given here.\citep{Pea99}

The average density $\rho$ is typically separated into the known energy forms: matter, radiation, and a cosmological constant. Then 
\begin{equation}
\rho = \rho_0 + { \rho_3 \over a^3} + {\rho_4 \over a^4} .
\end{equation}
Here, $\rho$ depends on $a$, and hence the time $t$, as the universe evolves. Classically, $\rho_0 c^2$ is associated with the cosmological constant (sometimes labeled $\Lambda$), $\rho_3 c^2$ is the present value of matter energy density, and $\rho_4 c^2$ is the present value of the radiation energy density. 

It is instructive to separate the energy density further into hypothesized energy forms. We expand $\rho$ in inverse powers of $a$ and write
\begin{equation}
\label{eq:rho}
\rho = \sum_{n=-\infty}^{\infty} \rho_n \, a^{-n} .
\end{equation}
The $\rho_n$ values on the right-hand side of Eq.~\eqref{eq:rho} remain fixed to their values at the scale factor $a=1$. 
It will be assumed that only positive values of $\rho_n$ and $a$ can exist. We define a critical density $\rho_c = 3 H^2/(8 \pi G)$ and let $\Omega = \rho/\rho_c$. We divide both sides of Eq.~\eqref{eq:rho} by $\rho_c$ at $a=1$ and obtain
\begin{equation} \label{OmegaSum}
\Omega = \sum_{n=-\infty}^{\infty} \Omega_n \,a^{-n} .
\end{equation}
$\Omega$ depends on $a$, and hence the time $t$, whereas all of the $\Omega_n$ values are fixed at the time when $a=1$. 

If we consider only the epoch of $a=1$, we have
\begin{equation} \label{generalizedf1}
\Omega_{\rm total} = \sum_{n=-\infty}^{\infty} \Omega_n .
\end{equation}
We rewrite this sum using popular labels such as 
\begin{eqnarray} \label{omegatotal}
\Omega_{\rm total} &= & \Omega_{\text{phantom energy}} 
+ \Omega_{\text{cosmological constant}} 
+ \Omega_{\text{domain walls}} \nonumber \\
&&{} \quad \Omega_{\text{cosmic strings}}
+ \Omega_{\rm matter} 
+ \Omega_{\rm radiation} 
+ \Omega_{\rm ultralight} .
\end{eqnarray}
These labels will be explained in detail in the following sections.

The curvature term in Eq.~(\ref{FriedmannOld}) can be written in terms of more familiar quantities. We divide Eq.~(\ref{FriedmannOld}) by $H^2$and rewrite Eq.~(\ref{FriedmannOld}) as
\begin{equation}
1 = \Omega - { k c^2 \over H^2 R^2 } .
\end{equation} 

Given these inputs, a more generalized Friedmann equation of energy can be written in a dimensionless form that explicitly incorporates all possible stable, static, isotropic energy form described by an integer $n$.  Equation~(\ref{FriedmannOld}) is divided by the square of the Hubble parameter when $a=1$, $H_0^2$, so that 
\begin{eqnarray} \label{friedmanna}
\Big(\frac{H}{H_0}\Big)^2 &=& \sum_{n=-\infty}^{\infty} \Omega_n \,a^{-n}
+ (1 - \Omega_{\rm total}) \,a^{-2} , \\
\noalign{\noindent or}
\label{friedmann1}
(H/H_0)^2 
&\approx & 
\sum_{n = -\infty}^{-1} \Omega_{\text{phantom energy}} \,a^{-n} + \Omega_{\text{cosmological constant}} \,a^0 
+ \Omega_{\text{domain walls}} \, a^{-1} \nonumber \\
&&{} + \Omega_{\text{cosmic strings}} \, a^{-2} 
+ (1 - \Omega_{\rm total}) \, a^{-2} + \Omega_{\rm matter} \, a^{-3} \nonumber \\
&&{}+ \Omega_{\rm radiation} \, a^{-4} + \sum_{n=5}^{\infty} \Omega_{\rm ultralight} \, a^{-n} . \label{fried}
\end{eqnarray}
Equation~\eqref{fried} can be written to highlight only present day observables by substituting $1/a = R_0/R = (1+z)/(1+z_0) = (1+z)$, where $z$ is the redshift and $z_0 = 0$ is the redshift at the epoch where $a=1$. 

\section{Attributes of the Friedmann Equations}

\subsection{The gravitational horizon}

Over what scale lengths do the Friedmann equations operate? The maximum scale length is referred to as the ``gravitational horizon." Energy outside of an object's gravitational horizon will have no gravitational effect on that object.\citep{Cal01} There is little observational evidence that limits the maximum size of the current gravitational horizon of our local universe. If the gravitational horizon operates like a light cone, it would be expanding at the speed of light.\citep{Nem05} The gravitational horizon of a point in space would be exactly the particle horizon for that point, encompassing the volume of space from which light or any relativistic particle could have come to that point since the universe started.\citep{Dav01} In a sense, we can only be gravitationally affected by energy forms that we can see.

Inside the gravitational horizon only anisotropic distributions of energy can cause local energy forms to gravitate toward any particular direction. Isotropic energy distributions that surround any particular point, no matter how distant, will not accelerate matter at that point with respect to the rest frame of the universe, as stated by Birkhoff's theorem.\citep{Jeb21}  Similarly, the expansion rate of the universe inside any given sphere surrounded only by isotropic energy is determined by applying the Friedmann equations to energy distributed inside the sphere.\citep{Gol04}

\subsection{The rest frame with respect to the universe}

In cosmology a ``rest frame with respect to the universe" or ``cosmic rest frame" is frequently cited. Such a rest frame might seem counter-intuitive, because special relativity has no preferred inertial frames. Every place in the universe has a definable rest frame, delineated by the energy in the universe itself. The universe rest frame at any point is the one frame where the average velocity of energy forms in the universe is zero. In this frame, for example, the cosmic microwave background photons will appear the same in all directions. It is usually assumed that all forms of energy are at rest on the average with respect to all other forms of energy. Note that were any form of energy to dominate the universe, the frame where this energy form is at rest would determine the rest frame for the universe. 

Our Earth is not so privileged so as to be at rest with respect to the cosmic microwave background frame. Analyses of the dipole of the microwave background seen from Earth shows that our Milky Way Galaxy is moving at about 600\,km/s with respect to this rest frame.\citep{Lin96} 

\subsection{Energy conservation}

What do the Friedmann equations mean in terms of energy conservation over any finite (non-local) scale? Very little. The Friedmann equations define only how the energy which is in the universe drives the expansion of the universe. As we have indicated, the energy density of any stable energy form $n$ goes as $a^{-n}$, decreasing as the universe expands for positive $n$. Because this energy dilution occurs everywhere in the universe, there is no place for the energy to go. Within the confines of the Friedmann equations, measured with respect to the rest frames of the universe, this energy is lost. 

We might find solace by demanding that single components of energy remain constant and conserve energy individually as the universe evolves. As will be seen in the following sections, examples where single components do not themselves change energy include normal matter, cosmic strings, and domain walls. The mass of a single electron, for example, does not change as the universe expands. In these cases the loss of energy density is still unexplained, but it can at least be completely attributed to the geometric expansion of the universe. 

There is at least one type of known energy which has components that change energy as the universe expands: radiation. Individual pieces of radiation, photons, for example, become redshifted and individually lose energy as the universe expands. 

More generally, the speed relative to the cosmic rest frame of any freely moving component of any form of energy will decay as the universe expands. This kinetic energy decrease related to the reduced speed does not go anywhere -- it just disappears from the universe and from the Friedmann equations. In general, a relativistic energy form has an integer number $n$ one integer higher than the same form of energy at rest.  A specific example of this generality is that radiation is the relativistic energy form of rest matter.

To better understand the kinetic energy loss, consider a particle moving at a relativistic speed $\gamma = 1/\sqrt{1 - v^2/c^2}$ with respect to the cosmic rest frame. Suppose that this particle was detected at speed $v$ by an observer at redshift $z$. What will be this particle's speed when it arrives locally at $z=0$? The quantity $\gamma v a$ is constant for anything moving with speed $v$ with respect to the cosmic rest frame.\citep{Lig75} In terms of the redshift, $\gamma v a = \gamma v (1+z)^{-1}$. In general, $\gamma v a$ is proportional to the momentum of the particle, but for relativistic particles, $v \approx c$ so that the energy of the particle is $c$ times the momentum. Therefore, for relativistic particles, the kinetic energy of any particle decays as $(1+z)$, just like radiation. As an example, it is straightforward to see that a particle detected with a highly relativistic velocity of $\gamma = 100$ at a redshift of $z=1$ will be detected to have a $\gamma \approx 50$ locally. 

It is interesting to estimate the loss rate of kinetic energy by all non-relativistic $n \approx 3$ matter in the present expanding universe. For non-relativistic matter, $\gamma \approx 1$, so that $\gamma v a \approx v a$ is approximately constant. The time derivative of this invariant gives $a dv/dt + v da/dt = 0$. Therefore $dv/dt = -(v/a) da/dt = -v H$, where $H=(1/a)(da/dt)$. 

For non-relativistic particles the kinetic energy density $\rho_K c^2 \approx (1/2) \rho v^2$, where $\rho=\rho_3 a^{-3}$ represents the particle density. Therefore the rate of change of the particle kinetic energy density is $d\rho_K c^2/dt = (1/2) v^2 d\rho/dt + \rho v dv/dt$. Note that $d\rho/dt = d (\rho_3 a^{-3})/dt = -3 \rho_3 a^{-4} da/dt = -3 \rho_3 a^{-3} H$. Combining these relations gives the rate of change of the kinetic energy as $d\rho_K c^2 /dt = (1/2) v^2 (-3 \rho_3 a^{-3} H) - \rho_3 a^{-3} v^2 H$. The first term gives the amount of kinetic power lost by the diluting mass density, and the second term gives the amount of kinetic power lost to particles slowing. When $a=1$ at the present epoch, these terms combine to give $d\rho_K c^2/ dt = (-5/2) v^2 \rho_3 H_0$. 

As we apply this analysis to the present universe, we will assume that all matter in the universe, including dark matter, has an effective speed relative to the cosmic rest frame of $v = 600$\,km/s, that matter makes up 30\% of the critical energy density, and that $H \approx H_0 \approx 70$\,km\,s$^{-1}$\,Mpc$^{-1}$ today.\citep{Lin96, Rie05} Therefore $\rho_c \approx 3 H_0^2/(8 \pi G) \approx 10^{-17}$\,kg\,km$^{-3}$ is the critical density today. Given these values, the average kinetic energy density disappearing from matter in the universe in one second is about $6 \times 10^{-33}$\,kg\,m$^{-1}$\,s$^{-3}$ if measured today. The energy of a single hydrogen atom is about $m_P c^2 \approx 1.5 \times 10^{-10}$\,kg\,m$^2$/s$^{2}$, which approximately corresponds to the loss of one hydrogen atom over the volume of the Earth every thirty seconds. 

At any one time in the universe all of the energy is typically constrained to be $\Omega_{\rm total}$. The statement $\Omega_{\rm total} = 1$ is a form of energy conservation, and implies that the local energy, no matter its form, is just enough to cause spacetime to be flat. A value of $\Omega_{\rm total} = 1$ is indicated by many present astronomical observations.\citep{Spe03} If $\Omega_{\rm total}$ is constrained to be exactly unity, it cannot be moved away from unity within the bounds of standard Friedmann cosmology, even if all of its constituent $\Omega_n$ values vary wildly with $z$. 

In Newtonian gravity the energy of a rising free body moves from kinetic to potential energy. Potential energy is a mathematical construct used to balance energy conservation equations. There is no equivalent potential energy for Friedmann cosmology. In general, energy is not conserved over large distances and times in general relativity.\citep{Wei02}

The general issue of conservation of energy is subtle in general relativity. Let's start with the time component of $T^{\alpha \beta}$, assuming that energy is conserved
\begin{equation} \label{energycon}
\nabla_{\mu} T^{\alpha \beta} = 0 .
\end{equation}
Because we have taken only the time component of $T^{\alpha \beta}$, Eq.~(\ref{energycon}) represents only conservation of energy. The remaining components yield momentum conservation along the three spatial directions. If we expand Eq.~(\ref{energycon}) using the rules of covariant differentiation, we obtain 
\begin{equation} \label{energycon1}
{ \partial T^{\alpha \beta} \over \partial x^{\mu} } + 
\Gamma^{\alpha}_{\mu \lambda} T^{\lambda \beta} + 
\Gamma^{\beta}_{\mu \lambda} T^{\alpha \lambda} = 0 .
\end{equation}

If a version of Eq.~(\ref{energycon1}) is considered locally, the Christoffel connections would vanish, leaving only partial derivatives with respect to time. Noether's theorem states that energy is time translation invariant.\citep{Pea99} A physical quantity is conserved only if the coordinates that it generates are missing from the Lagrangian. Because energy is the generator of infinitesimal time translations, the appearance of time in the Lagrangian indicates that energy is not conserved over finite times and finite distance scales. Nevertheless, energy is considered to be conserved on infinitesimal scales (``locally") in general relativity, over which curvature changes are not important. 

\subsection{Perfect fluids and gravitational pressure}

The Friedmann equations typically quantify the cosmological evolution of perfect fluids. Perfect fluids are characterized by only two variables: the energy density $\rho$ and isotropic pressure $P$. These two variables can be isolated in an effective modified Poisson equation for gravity from general relativity in the weak field limit so that\citep{Pea99}
\begin{equation} \label{poisson}
\nabla^2 \phi = 4 \pi G (\rho + 3 P/c^2) ,
\end{equation}
where $\phi$ is the Newtonian potential in the limit of weak field gravity. The gravitational influence of pressure has no Newtonian analogue.

In general, the equation of state of a perfect fluid is given by $w = P/(\rho c^2)$ so that $\rho \propto a^{-3(1+w)}$.\citep{Hut01} If energy is conserved locally for a perfect fluid, $n= 3w + 3$ and $w = n/3 - 1$. The terms in the Friedmann equations with $w<0$ ($n<3$) have repulsive (negative) gravitational pressure. Locally conserved perfect fluids will be assumed by default for the rest of this paper. 

For a universe dominated by a single form of perfect fluid energy where $w$ is unchanging, $c_s^2 = \partial P/\partial \rho = w$, where $c_s$ is the sound speed in the perfect fluid. When $w < 0$ ($n < 3$), the sound speed becomes imaginary, causing instabilities. Instabilities in these cosmological ``fluids" may exist only on small scales and may not affect the large scale nature of the universe. Therefore, it is common to assume that all perfect fluid energy forms act homogeneously on cosmological scales. 


Another phenomenological problem occurs for perfect fluids with $w < - 1$ ($n < 0$) and $w>1$ ($n>6$). Energy forms with this property have a sound speed greater than the speed of light, possibly making them unphysical.\citep{Gon03}

Perfect fluids are not the only hypothesized forms of energy. A simple alternative is the ``Chaplygin gas" which has $P = -A/\rho$, where $A$ is a positive constant.\citep{Dev03} Given a positive $\rho$, a Chaplygin gas would also have a positive sound speed. This type of fluid will not be considered further.

\subsection{The ``other" Friedmann equation: The Friedmann equation of acceleration}

The Friedmann equation of energy, Eq.~(\ref{friedmanna}), is the time-like part of a larger four-vector with three spatial components. Each of the three space-like components is assumed identical because all energy forms are considered to be spatially isotropic. The single remaining unique space-like component is referred to as the Friedmann equation of acceleration. Both Friedmann equations are derived from the same set of assumptions. The Friedmann equation of acceleration is less cited and more difficult to treat because it is a second-order differential equation. Second order indicates the accelerative nature of this equation. The Friedmann equation of acceleration is typically written in the form\citep{Pea99} 
\begin{equation} \label{FAinit}
{\ddot a \over a} = -{4 \pi G \over 3} (\rho + 3 P/c^2).
\end{equation}

Like the Friedmann equation of energy, the Friedmann equation of acceleration can also be written in a dimensionless fashion that explicitly incorporates all integer $n$ energy forms. To do so we must assume an equation of state that relates the pressure to the density. The most commonly used equation of state is the perfect fluid relation where energy is locally conserved. In terms of $w$, the expanded Friedmann equation of acceleration becomes
\begin{equation}
\label{eq:a}
{ \ddot{a} \over a} = - \left( {H_0^2 \over 2} \right)
\left( {8 \pi G \rho \over 3 H_0^2} \right)
(1 + 3w) .
\end{equation}
As with the Friedmann equation of energy, we define a critical density $\rho_c = 3 H^2/(8 \pi G)$ and let $\Omega = \rho/\rho_c$.  Equation~\eqref{eq:a} can also written in terms of the present epoch where $a=1$, $H=H_0$, $n=3w+3$, and $\Omega_n$'s are fixed at their $a=1$ values. Then, using Eq.~(\ref{OmegaSum}), 
\begin{equation} \label{generalizedf2}
{ \ddot{a} \over a} = H_0^2 \sum_{n=-\infty}^{\infty} 
\left( 1 - { n \over 2} \right) \Omega_n \,a^{-n} .
\end{equation}

Note that this Friedmann equation of acceleration can also be obtained by taking the time derivative of the Friedmann equation of energy. In this sense, if $P$ and $\rho$ are related by a perfect fluid equation of state and local energy is conserved, the two Friedmann equations are not independent. 

An expanded Friedmann equation of acceleration written with labels reads 
\begin{eqnarray} \label{friedmann1b}
{\ddot{a} \over H_0^2 a} 
& \approx & \sum_{n=-\infty}^{-1} (1 - {n \over 2}) 
\Omega_{\text{phantom energy}} \, a^{-n} + \Omega_{\text{ cosmological constant}} \,a^0
+ {1 \over 2} \Omega_{\text{domain walls}} \,a^{-1} \nonumber \\
&&{}- {1 \over 2} \Omega_{\rm matter} \,a^{-3} 
- \Omega_{\rm radiation} \, a^{-4} + \sum_{n=5}^{\infty} (1 - {n \over 2}) \Omega_{\rm ultralight} \,a^{-n} .
\end{eqnarray}
Both the $n=2$ cosmic string term and the $n=2$ curvature term, tracked by the $(1 - \Omega_{\rm total})$ term in Eq.~(\ref{friedmann1}), are absent because they do not alter the magnitude of the acceleration of the universe.

\subsection{Minimally simplified Friedmann equations}

We can insight by rewriting the Friedmann equations only in terms of how the scale factor $a$ changes with time when $\Omega_{\rm total} = 1$ and one form of energy $n$ (or $w=n/3 - 1$) dominates. Then, the Friedmann equation of energy, specifically Eq.~(\ref{friedmanna}), can be combined with $H = ({\dot a}/a)$ to yield
\begin{subequations}
\label{FEminall}
\begin{align} \label{FEmin}
{\dot a} &= H_0 a^{1-n/2} , \\
\noalign{\noindent and}
\label{FAmin}
{\ddot a} &= H_0^2 (1 - n/2) a^{1 - n} .
\end{align}
\end{subequations}
Equation~\eqref{FEminall} can be written as
\begin{subequations}
\begin{align} \label{FEminw}
{\dot a} &= H_0 a^{- {3 \over 2} w - {1 \over 2}} ,\\
\noalign{\noindent and}
{\ddot a} &= H_0^2 \left( - {3 \over 2} w - {1 \over 2} \right) a^{-3w - 2} .
\end{align}
\end{subequations}

\subsection{Minimally simplified Friedmann solutions}

Intuition about the evolution of Friedmann universes can be gained by solving the Friedmann equations for $a(t)$ when $\Omega_{\rm total} = 1$ and one form of energy $n$ (or $w=n/3 - 1$) dominates. The solutions can be generalized when $w>-1$ ($n>0$). The Friedmann equation of energy, Eq.~(\ref{FEmin}) can be written as
\begin{equation} \label{generalequation}
a^{(n-2)/2}\,da = H_0 \,dt .
\end{equation}
Integration of Eq.\eqref{generalequation} gives 
\begin{equation}
\left( { 2 \over n } \right) a^{n/2} 
= H_0 t + k ,
\end{equation}
where $k$ is an integration constant. For $n > 0$ and $a = 0$ when $t = 0$ we have $k = 0$. Therefore,
\begin{align} \label{generalsolution}
a &= \left( { n \over 2} \right)^{2/n} (H_0 t)^{2/n} ,\\
\noalign{\noindent or in terms of $w$} 
\label{GSw}
a &= \left( {3 \over 2} w + {3 \over 2} \right)^{2 \over 3w+3}
(H_0 t)^{2 \over 3w+3} .
\end{align}
Solutions to Friedmann's equations for an universe where multiple energy forms are simultaneously important are given in Ref.~\onlinecite{Lak06}.

\section{Energy Forms in the Friedmann Equations}

Energy can take any number of forms -- it may never be known how many. Energy is usually referred to by the form it takes.  Forms of energy can be classified by how they gravitationally affect the universe in the Friedmann equations. Stable, integer energy forms change their energy density relative to the cosmic rest frames only as an integer power of the universe scale factor: $\rho \propto a^{-n}$. Static matter and isotropic radiation are examples of stable, integer energy forms. The number $n$ of an energy form can be interpreted as the number of effective spatial dimensions in which a unit of an energy form is confined. Energy forms with $w < 0$ ($n < 3$) are typically referred to as ``dark energy." A list of all known and hypothesized perfect fluid energy forms is given in Table~\ref{tab1}. How these stable energy forms interact are extrapolated from Ref.~\onlinecite{Vil81a} and are summarized in Table~\ref{tab2}.

\subsection{Phantom energy: $w < -1$}

Forms of energy in the Friedmann equations with both negative $w$ and $n$ are called phantom energy. Phantom energy is unconfined in any spatial direction and pervades the entire universe. Any positive $\Omega_{\text{ phantom energy}}$ that multiplies any term with $w < -1$ ($n<0$) increases in magnitude as the universe expands. If no other phantom energy terms exist, and if the given phantom energy does not undergo a phase transition, then it will eventually grow from any miniscule density to dominate the energy and expansion dynamics of the universe.\citep{Cal03} One reason that phantom energy is controversial is because its formal sound speed, $c_s = c \sqrt{w}$, has a magnitude greater than the speed of light. 

To see how the universe evolves with time in $\Omega_{\rm total} = 1$ cosmologies dominated by phantom energy, we cannot use the single-term general solution Eq.~(\ref{generalsolution}) because it applies only to $w > -1$ ($n>0$). However, it is straightforward to integrate Eq.~(\ref{FEmin}), which becomes for $w < -1$, 
\begin{equation}
\left( {2 \over n} \right) a^{n/2} 
- \left( {2 \over n} \right) = H_0 t .
\end{equation}
Note that $t$ is measured from the time that the $w<-1$ ($n<0$) phantom energy term begins to dominate the expansion of the universe. Then
\begin{align}
a &= \left[ { 1 \over 1 - (-n/2) H_0 t} \right] ^{2/(-n)} ,\\
\noalign{\noindent when written in terms of $w$ becomes}
a &= \left[ 1 \over 1 - {{ - 3 w - 3} \over 2} H_0 t \right]^{2 \over -3 w - 3} .
\end{align}

The universe scale factor $a$ goes to infinity in a finite amount of time, when the denominator approaches zero. That is the time of the ``Big Rip."\citep{Cal03} For very negative values of $3w$ and $n$, the phantom energy rips apart the universe quickly after dominating the expansion rate, while for values of $3w$ near $-1$ ($n$ near zero), the Big Rip will occur far into the future.

\subsection{Volume energy: Cosmological constant: $w=-1$}

The fundamental type of energy that evolves as $w=-1$ ($n=0$) in the Friedmann equations has energy distributed uniformly in space and is known as the cosmological constant or $\Lambda$. This form of energy is unconfined in any spatial direction. Note that not just any volume of energy would evolve as $w=-1$. Cosmological constant energy must remain isotropic with respect to all rest frames in the universe, even as the universe expands. We cannot make $w=-1$ energy by simply spreading $w=0$ ($n=3$) particles uniformly because these particles would either expand with the universe and so change their average energy density, or move increasingly rapidly with respect to the expanding universe, with the possible exception of a single point.

Energy forms with $w=-1$ have been hypothesized as important components of the universe as far back as Einstein, whose original intent was to create a repulsive gravity component that keeps a static universe of $w=0$ matter particles from collapsing. Since then, a cosmological constant has been suggested several times to account for possible anomalies in cosmological data.\citep{Car01} Such introductions were often met with skepticism. Present supernova data indicate that standard candle supernovae appear less bright than expected with only $w=0$ matter dominating the universe. These data are well fit with a significant high-density $w=-1$ cosmological constant term.\citep{Per99, Rie98} Additionally, current analyses of cosmic microwave background data indicate a flat $\Omega_{\rm total} = 1$ universe has only a $w=0$ component of about 30\%, consistent with $w=-1$ making up the rest.\cite{Spe03} Both of these measurements are also consistent with current galaxy clustering estimations of $w$ and have led to a concordance model where the $w=-1$ component makes up about 70\% of the energy of the universe.\citep{Teg01}

Cosmological constant $w=-1$ energy has gravitationally repulsive pressure. Although the energy density of $w=-1$ matter is positive, and is gravitationally attractive, Eq.~(\ref{FAinit}) shows that the repulsive pressure is three times greater in magnitude.  This repulsive pressure effectively accelerates the expansion of the universe. A summary of how $w=-1$ energy interacts with other forms of energy is given in Table~\ref{tab2}. 

Suppose that $w=-1$ dark energy could somehow be completely contained inside non-interacting boxes of constant size. Would these boxes still cosmologically evolve as $w=-1$? The answer is no. As the universe expands, each box would still contain the same amount of $w=-1$ energy, but the energy density between the boxes would be zero. As the boxes became further apart, the number density of boxes would decrease like $w=0$. Therefore, all of the boxes together would evolve as $w=0$. 

Similarly, if a small box containing cosmological constant energy dropped near the Earth, it would fall just as if the box were filled with a small amount of $w=0$ particles. This behavior is one realization of the equivalence principle. 

It is not possible for pervasive $w=-1$ energy to move with respect to the universe.\citep{Mar06} Conversely, it is also not possible for particles of $w=0$ matter to move with respect to pervasive $w=-1$ energy. Therefore, the motion of $w=0$ matter in a universe with $w=-1$ energy will yield no drag or accelerative aberration force. 

The Friedmann equation of energy for the case that a $w=-1$ ($n=0$) energy form solely dominates an $\Omega_{\rm total} = 1$ universe can be solved analytically. The solution is not given by Eq.~(\ref{GSw}) which applies only when $w>-1$. The solution is found by directly integrating Eq.~(\ref{FEminw}) which yields 
\begin{equation}
\ln a = H_0 t + \ln a_0 ,
\end{equation}
where $t$ can be considered the time since $w=-1$ energy began to dominate the expansion, when $a = a_0$. The solution is
\begin{equation} \label{dSsolution}
a = a_0 \, e^{H_0 t} .
\end{equation}
Equation~\eqref{dSsolution} is a classic solution where a cosmological constant drives an exponentially accelerating universe. The universe is said to be in a de-Sitter phase, and the Hubble parameter $H$ is static at $H_0$. Such a phase is hypothesized to have dominated the early universe in a phase called inflation.\citep{Gut81} Our universe appears to be slowly re-entering into another such phase now. 

\subsection{Sheet energy: Domain walls: $w=-2/3$}

A fundamental type of energy which evolves as $w=-2/3$ ($n=1$) in the Friedmann equations has energy confined to thin sheets. The term ``sheet" is used to mean that the thickness of the contained energy in one dimension is small compared to the gravitational horizon size. For a $w=-2/3$ universe, the averaged cosmic density distribution for all sheets must be isotropic in all cosmic rest frames and at all times of an evolving universe. Note that not just any sheet of energy would dilute as $w=-2/3$. A sheet of energy that acts as $w=-2/3$ is by definition not moving in the direction perpendicular to its area, with respect to the rest frames of surrounding universe, at every point along its entire surface, at all times. The sheet also retains its surface energy density as the universe expands. This surface energy invariance would not be possible, for example, for a sheet of $w=0$ ($n=3$) particles. A sheet that started in the rest frame of the universe at all points on the sheet would quickly dilute its average surface energy density if it expanded with the universe. If the sheet of $w=0$ particles did not expand with the universe, only a single point in the sheet would be able to remain in the rest frame of the universe. Points on the sheet far from this center would quickly be seen to moving relativistically, relative to local matter, as the universe expanded. 

Therefore, $w=-2/3$ sheet energy is intrinsically different than a sheet of uniformly spread $w=0$ particles, and any other energy form $w$. Such a $w=-2/3$ form of energy is called a (cosmological) {\it domain wall}.\citep{Kib76} A domain wall might arise as a topological defect that occurred during a phase transition in the early universe. Note that because the surface energy density in the $w=-2/3$ sheet remains constant as the universe expands, the sheet does not stretch to become thinner.

Were domain walls to form before the last epoch of inflation, they would likely have been diluted by inflation to be too sparse to be cosmologically observable or important today. No credible claim for the detection of a cosmological domain wall has been made. Nevertheless, attempts have been made to describe the expansion history of our universe with a network of domain walls.\citep{Buc89, Con04} 

The concept of domain walls is found in other branches of physics. A definition of a (non-cosmological) domain wall is a region between two volumes that formed slightly differently. A common example of a non-cosmological domain wall can be seen (and heard) in a glass of ice when water is poured into it. Many times, an ice-cube will ``crack" and show internal sheets that are defects in the pure ice. These cracks have likely formed among (several) domain walls in the ice.\cite{ice}.

If a person were to go up to a flat, static, cosmological domain wall, that person would feel gravitational repulsion from it.\citep{Vil81a} Although the positive energy density in a domain wall is gravitationally attractive, the stronger negative pressure will result in a net gravitational repulsion, as indicated in Table~\ref{tab2}. 

Domain walls would act differently in a cosmological sense if they were curved.\citep{Vil81a, Vil81b} Assume that it is possible that domain walls could be curved into balls.\citep{Vil81a} This situation is similar to containing any other form of energy in a $w=0$ box. As the universe expands, each domain ball would still contain the same amount of energy, but the energy between the balls would be zero. The number density of the balls would therefore decrease like $w=0$ particles as the universe expanded. Therefore, all of the balls together would evolve in the Friedmann equations as $w=0$. 

If a small box containing a domain ball were dropped near the Earth, it would fall as if the box were filled with any small amount of $w=0$ particles.\citep{Vil81a} That is one aspect of the equivalence principle. 

An intermediate case between domain walls and domain balls occurs when a domain wall obtains curvature on the scale of the gravitational horizon size. This sheet curvature will create an energy form between domain walls and domain balls, with an effective value of $w$ in the range $-2/3 < w < 0$. 

A domain wall moving perpendicular to its plane will increase $w$ so that $-2/3 < w < -1/3$. The increase in $w$ derives from the same physics that causes photons to lose energy -- the wall continually moves into frames where it has less speed relative to the cosmic standards of rest. When moving relativistically, the value of $w$ of a domain wall will approach $w \approx -1/3$. 

If domain walls dominate the Friedmann equations, $H^2 = \Omega_1 H_0^2 a^{-1}$ so that $({\dot a}/a)^2 \approx a^{-1}$, and hence $a \approx t^2$.

If a domain wall energy sheet were rolled up into a static, straight, hollow tube so that the radius of the tube is small compared to the gravitational horizon, but the length of the tube is long compared to the gravitational horizon, the resulting ``domain tube" would act like cosmological $w=-1/3$ ``string" matter.

\subsection{Line energy: Cosmic strings: $w=-1/3$}

A fundamental type of energy that evolves as $w=-1/3$ ($n=2$) in the Friedmann equations has energy confined to lines. ``Line" is used to mean that the confinement radius of the energy in two dimensions is small compared to the gravitational horizon size. Not just any line of energy would dilute as $w = -1/3$ energy. For $w=-1/3$ the line density distribution must be isotropic in all cosmic rest frames and at all times of an evolving universe. A line of energy that acts as $w=-1/3$ is by definition not moving in the direction perpendicular to its length with respect to the rest frames of the universe, at every point along its length at all times. The line also retains its one-dimensional energy density as the universe expands. This energy density invariance would not be possible for a line of $w=0$ ($n=3$) particles. Such a particle line that started in the rest frame of the universe at all points in the line would quickly dilute in energy density as $w=0$ if it expanded with the universe. If the $w=0$ line did not expand with the universe, only a single point in the line would be able to remain in the rest frame of the universe -- points on the line far from this ``center" would quickly be seen to moving relativistically, relative to local matter, as the universe expanded. 

Therefore, a $w=-1/3$ line energy is intrinsically different from a line of uniformly spread $w=0$ particles, and any other energy form $w$. Such a $w=-1/3$ energy form is called a {\it cosmic string}.\citep{Vil81b} A cosmic string might arise as a topological defect that occurred during a phase transition in the early universe.\citep{Kib76} Note that because the linear energy density in the $w=-1/3$ string remains constant as the universe expands, the string does not ``stretch" to become thinner.

If cosmic strings were to form before the last epoch of inflation, they would likely have been diluted by inflation and be too sparse to be cosmologically observable or important today. No enduring claim for the detection of a cosmic string has been made. 

The concept of topological defect strings is not foreign to other branches of physics. A general definition of a topological string is a region between two areas that formed slightly differently. An example of non-cosmological topological strings can be seen on a pond when it freezes to ice. Many times, sheets of surface ice meet along wandering boundaries that can be considered strings. 

If a person were to go up to a straight, static, cosmic string, that person would feel no gravitational attraction toward or away from it.\citep{Vil81a} This neutral behavior is indicated in Table II. Although the positive energy density in a cosmic string is gravitationally attractive, the negative pressure is gravitationally repulsive with the same magnitude, so that the combined result yields no local gravitational effect. 

How would cosmic strings act if they were curved into closed loops? This situation is similar to containing any other form of energy in a box. As the universe expands, each cosmic string loop would still contain the same amount of energy, but the energy between the loops would be zero. The number density of loops would therefore decrease like $w=0$ matter particles. Therefore, all of the loops together would evolve as $w=0$ particles as the universe expanded. 

Similarly, if a small box completely containing a cosmic string loop were dropped near the Earth, it would fall as if the box were filled with any small amount of $w=0$ particle energy, as dictated by the equivalence principle. 

A string with significant curvature with a scale less than the gravitational horizon will act like an energy form intermediate between no-curvature, long cosmic strings, and high-curvature, short string loops. The curved string would have an effective equation of state parameter $-1/3 < w < 0$. A cosmic string moving perpendicular to its length will increase $w$ so that $-1/3 < w < 0$. The increase in $w$ derives from the same physics that causes photons to lose energy -- the string continually moves into frames where it has less speed relative to cosmic standards of rest.
When moving relativistically, the $w$ of a cosmic string would actually approach $w \approx 0$ ($n=3$). 

When cosmic strings dominate the Friedmann equations, $H^2 = \Omega_2 H_0^2 a^{-2}$ so that $({\dot a}/a) \approx a^{-1}$. If we solve for $a$ as a function of time, it is straightforward to see that $a \approx t$.

\subsection{Point energy: Matter: $w=0$}

A fundamental type of energy that evolves as $w=0$ ($n=3$) in the Friedmann equations has energy confined to a point. ``Point" is used to mean that the confinement radius of the energy in all three spatial dimensions is small compared to the gravitational horizon size. Examples of $w=0$ particles include the reader, baryonic matter, dark matter, massive fundamental particles, and topological defects such as magnetic monopoles.\citep{Pre92, Blu84} Conglomerates of $w=0$ particles such as heavy nuclei, molecules, and black holes also act cosmologically as $w=0$. These particles may be held together by a fundamental force, such as the strong nuclear force, electromagnetism, or even gravity. 

As we have indicated $w=0$ components can be made by containing other forms of energy. Volume energy $w=-1$ could act like $w=0$ energy if itwere somehow able to be confined to finite three-dimensional boxes. Domain wall $w=-2/3$ energy could act like $w=0$ point energy if the domain wall curled up into a compact object. Cosmic string $w=-1/3$ energy could act like $w=0$ point energy were the string curled up into a compact object. Even $w=1/3$ radiation could be confined to a small mirrored box and act as $w=0$ energy. 

If we were to go near a static $w=0$ energy component, we would feel the classic gravitational attraction to it that Newton discovered over 300 years ago: $F = G M m/r^2$. Static forms of energy with $w=0$ exhibit no gravitational pressure.

A particle moving in any spatial direction will increase $w$ so that $0 < w < 1/3$. When moving relativistically, the $w$ of a particle would actually approach $w \approx 1/3$ and be considered a form of radiation. 

It is interesting to wonder what the value of $w$ is for an average baryonic particle in the universe today. Suppose that the average speed of matter in our universe relative to the cosmic rest frame is $v$. The momentum $p$ of a particle moving with speed $v$ is $p = m_0 v/\sqrt{1 - v^2/c^2} = m_0 v \gamma$. Here $m_0$ is the rest mass of a particle with rest energy $E_0 = m_0 c^2$. Averaged over all three spatial dimensions, $P = p c^3 /3$ so that $w = P/(\rho c^2) = p c/(3 E_0) = m_0 v c \gamma/(3 m_0 c^2) = (v/3c)\gamma$. Given that our Galaxy's measured present speed of 600\,km/s with respect to the cosmic rest frame is typical of all particles in the universe, we have $w \approx 6.7 \times 10^{-7}$. In term of $n$ we use $\Delta n = 3 \Delta w = 2 \times 10^{-6}$, so that for the average particle in the present universe, $n \approx 3.000002$.

When particles dominate the Friedmann equations, then $H^2 = \Omega_3 H_0^2 a^{-3}$ so that $({\dot a}/a) \approx a^{-3/2}$ and $a \approx t^{2/3}$.

\subsection{Radiation: $w=1/3$}

A fundamental type of energy that evolves as $w=1/3$ ($n=4$) in the Friedmann equations has energy confined not only in all three spatial dimensions but moving relativistically with respect to its local rest frame in the universe. This familiar energy form is known as radiation. The most familiar example is electromagnetic radiation, but other commonly discussed examples include gravitational radiation and neutrinos. 

Theoretically, any $w=0$ particle could be accelerated to near the theoretical speed limit of the speed of light and subsequently act as effective $w = 1/3$ radiation in the cosmological sense of the Friedmann equations. Also, $w=1/3$ radiation components can be made from any energy form that can be spatially confined in three dimensions and boosted to a relativistic speed.

Does any radiation have $w=1/3$ exactly? We take the didactic view in this section that all radiation has the some amount of particle rest mass, even if it is minuscule, has no aggregate cosmological significance, and no reasonable possibility of being measured. Therefore, just as particle energy typically has a speed and so a $w$ slightly greater than zero, radiation energy might have some rest mass and so a $w$ slightly smaller than $1/3$. The present rest mass limits on radiation include the photon\citep{Luo03} at $1.2 \times 10^{-54}$\,kg, the neutrino\citep{Elg06} at about $1.8 \times 10^{-36}$\,kg, and gravitational radiation\citep{Ger98} at $4.5 \times 10^{-69}$\,kg. 


The gravitational pressure of $w=1/3$ radiation is positive. Because of gravitational pressure, radiation gravitates twice as strongly as $w=0$ matter energy with the same average energy density. This difference caused radiation dominated epochs of our early universe to expand more slowly than subsequent matter and cosmological constant epochs.

Suppose all radiation could be completely contained inside static non-interacting boxes. Would it still cosmologically evolve as $w=1/3$ radiation? The answer is no. As the universe expands, each box would still contain the same amount of radiation energy as measured inside each box. There is no impetus for the trapped radiation to lose energy as the universe expands. The energy density between the boxes would remain zero. Therefore, the number density of the boxes would decrease as the universe expands as the number of spatial dimensions $n=3$. All of the boxes together would therefore evolve as $w=0$ particle energy. 

If a small box containing radiation were dropped near the Earth, it would fall according to the equivalence principle as if the box were filled with $w=0$ energy.\citep{Kri90} 

When radiation dominates the Friedmann equations, then $H^2 = \Omega_4 H_0^2 a^{-4}$ so that $({\dot a}/a) \approx a^{-2}$, and $a \approx t^{1/2}$.

\subsection{Ultralight: $w>1/3$}

Are there any other types of energy in the universe? The only remaining realm for stable energy forms have $w>1/3$ ($n>4$). This specific possibility was discussed recently, and such energy forms were referred to as {\it ultralight}.\citep{Nem07} The term ultralight energy contrasts with dark energy as being beyond light in terms of attractive gravitational pressure. By analogy, ultralight is to light what ultraviolet light is to violet light.

The idea of energy evolving with an effective $w>1/3$ has been hypothesized in the context of time-varying scalar fields.\citep{Kam90} An epoch with $w=1$ might exist if a scalar field were to slow the expansion rate of the universe as it exited inflation.\citep{Spo93, Joy98} A contracting universe dominated by a scalar field with effective $w > 1$ would create homogeneity similar to an expanding, inflating universe.\citep{Ste05} However, the stable ultralight energy forms discussed in this section are not scalar fields. 

Forms of ultralight with $w>1$ might be unphysical because their formal sound speed $c_s = c \sqrt{w}$ is greater than the speed of light. Only ultralight energy form with $1/3 < w < 1$ would have a sound speed less than $c$. 

A particularly mundane form of ultralight energy has $w=2/3$. All of the known stable energy forms and all of the energy forms we have described have integer powers of $3w$ and $n$. The next in the linear integer progression after $n=4$ radiation is $n=5$ ultralight. The latter would dilute cosmologically as $a^5$, also an integer power of $a$. Furthermore, $w=2/3$ ultralight can be a stable perfect fluid because its sound speed $c_s$ is neither imaginary nor greater than $c$. Therefore, the $w=2/3$ ultralight energy form will be highlighted here as of particular interesting. 

Why would any energy form act with $w>1/3$? Although there are clear dimensional paths to understanding most other integer $3w$ forms, there is no such clear path for understanding ultralight. Ultralight might respond, at least in part, to other spatial dimensions, such as those spatial dimensions hypothesized in string theory\citep{Wit95} or Randall-Sundstrum cosmologies.\citep{Ran99} If so, such forms of energy might act only as lower $w$ energy forms as these dimensions became cosmologically unimportant. Alternatively, ultralight might somehow incorporate multiple sensitivities to the time dimension. 

The gravitational pressure of $w>1/3$ ultralight is attractive, so that $w>1/3$ ultralight cosmologically gravitates more strongly than any other form of energy. 

If a small box containing and confining ultralight energy were dropped near the Earth, the equivalence principle demands that it fall just as if the box were filled with any small amount of $w=0$ particle energy. 
If $w=2/3$ ultralight dominates the Friedmann equations, then $H^2 = \Omega_5 H_0^2 a^{-5}$, so that $({\dot a}/a) \approx a^{-5/2}$. Therefore, as the universe expands, ultralight dilutes even faster than light. It is straightforward to see that $a \approx t^{2/5}$. 

Ultralight is not considered to be a significant contributor to the energy budget of the universe today, although ultralight might have affected the universe in the distant past. The reason derives from the high attractive gravitational pressure of ultralight which would slow the expansion of the universe, and hence extend the duration of epochs when ultralight dominated. The strongest present limit on $w=2/3$ ultralight comes from primordial nucleosynthesis results.\citep{Nem07} If nucleosynthesis were dominated by $w = 2/3$ ultralight, the longer duration would create a higher fraction of heavy elements than detected today. Given that $\Omega_{\rm radiation} \approx 2 \times 10^{-5}$ in today's universe,\citep{Pea99}, $\Omega_{\rm ultralight} < 10^{-11}$ today.\citep{Nem07}

\subsection{More complicated forms of energy}

In addition to energy forms that change the energy density only as a power law of the universe scale factor $a$, other forms of energy might exist. One set of possibilities includes unstable topological defects of higher order than monopoles such as textures.\citep{Dav87, Tur89} Another set of possibilities involves energetic scalar fields such as the hypothesized field that fueled inflation, vector fields such as the electromagnetic field, and even tensor fields like the general relativistic formulation of gravity. A simple such field is a quantum scalar field $\phi$. Such fields may respond to a potential energy function $V(\phi)$ which may be driven on its own time scale and not be simply related to the universe scale factor. If invoked to explain dark energy observations in cosmology, a scalar field may be called {\it quintessence}.\citep{Pee88,Zla99} 

The Friedmann equation of energy dominated by such a scalar field $\phi$ is written as\citep{Pee88} 
\begin{equation}
H^2 = (1/2) {\dot \phi}^2 + (1/2) {\nabla \phi}^2 + V(\phi) .
\end{equation}
These fields typically define an equation of state such that 
\begin{equation}
w = { { {1 \over 2} {\dot \phi}^2 - V(\phi) } \over
{ {1 \over 2} {\dot \phi}^2 + V(\phi) } }.
\label{thiseq}
\end{equation}
Although $V$s exist that make $3w$ an integer, such cases are exceptional. From Eq.~\eqref{thiseq} any evolving scalar field for which $V(\phi) \ll {1 \over 2} {\dot \phi}^2$ for any phase will effectively have $w=1$ during that phase.\citep{Kam90} Alternatively, if $V(\phi) \gg {1 \over 2} {\dot \phi}^2$, then $w=-1$ for any finite potential $V$. The free parameters allowed by these cosmological fields give these fields the potential of being significantly more complicated than the ``stable" fields considered in other sections. They will not be considered further here. 

\section{Changes Between Forms of Energy}

\subsection{Slow changes}

Stable energy forms do not always stay in the same form. Changes between stable energy forms that occur on a time scale comparable to the total expansion time of the universe are referred to as slow. An example that has been discussed frequently in this context is the continuous cosmological slowing of energy moving relative to the cosmic rest frame as the universe expands. Given a monotonically expanding universe, even radiation anchored by an arbitrarily small rest mass will eventually decelerate to appear nearly at rest with respect to the cosmic rest frames, so that even this $w \approx 1/3$ radiation will eventually become $w \approx 0$ matter. 

How long will it be before today's $w \approx 1/3$ photons morph into $w \approx 0$ particles nearly at rest with respect to the expanding universe? The time depends strongly on the unknown rest mass of the photon and the possibly unknowable expansion trajectory of the future universe. Nevertheless, a very approximate lower mass limit can be estimated assuming that the present expansion rate applies indefinitely into the future and that photons have a rest mass right at the present experimental limit. 

For photons, $m_0 \approx 10^{-54}$\,kg is a recent upper limit of the rest mass.\citep{Luo03} This mass corresponds to a rest energy of $E_0 = m_0 c^2$ of about $10^{-65}$\,kg \,m$^{2}$/s. Let's assume that all photons have this rest mass. The total energy of a visible photon is about $E = 10^{-18}$\,kg\,m$^{-2}$\,s$^{-1}$. Remember that $E = E_0 \gamma$. Therefore, $\gamma = E/E_0 \approx 10^{47}$. 
If we solve for $v$, we find that these photons are going about $(1 - 10^{-94}) c$, where $c$ is the fastest possible speed. 

Recall the cosmological relation for matter moving at speed $v$ relative to the cosmic rest frame: $v \gamma = v_a \gamma_a a$, where the right-hand side quantities are measured in the cosmic rest frame when the universe has expansion factor $a$; the left-hand side quantities are measured in the cosmic rest frame when $a=1$. The relative expansion factor needed to bring photons to near their rest mass is therefore $a \approx \gamma \approx 10^{47}$. Specifically, this expansion factor will bring photons to a speed where $(v_a/c) \gamma_a \approx 1$, so that $v_a = c/\sqrt{2}$. 


Given a de Sitter cosmology where $w=-1$ energy dominates and $H = H_0$ remains constant into the future, the corresponding time this photon deceleration would take is given by Eq.~(\ref{dSsolution}) so that $a/a_0 = a/1 = e^{H_0 t}$. Therefore, $t = (\ln a)/H_0 \approx 108/H_0$ or about 108 ``Hubble times." Given that $1/H_0 \approx 1/(70$\,km\,s$^{-1}$\,Mpc$^{-1}) \approx 1.4 \times 10^{10}$ years, photons may become non-relativistic in about 1.5 trillion years in the future. 

Other methods exist where one energy form can slowly morph into another. An example is an energy form that slowly becomes confined as the universe evolves. Consider cosmic strings as an example. If $w=-1/3$ static cosmic strings were slowly increasing their curvature scale length relative to the scale length of the gravitational horizon, these strings might morph into loops wholly contained within the gravitational horizon, and hence act like $w=0$ matter particles as far as the Friedmann equations are concerned. The same logic holds for domain walls. Conversely, the gravitational horizon might expand from where the curvature of strings and walls were effectively negligible, to where strings and walls are effectively curved, yielding the same effect. 

\subsection{Fast changes}

Stable energy may undergo a rapid change of form. Possibly the most familiar transitions between energy forms occurs for energy shifts between $w = 0$ matter and $w = 1/3$ radiation. Examples include atomic, molecular, collisional, and nuclear transitions between $w=0$ matter particles which result in the emission of $w=1/3$ radiation such as photons. Conversely, $w=1/3$ radiation such as photons could collide and create $w=0$ matter.
When considering a rapid transition between energy forms in general, the magnitude of the energy densities is equivalent and do not depend on the magnitude or sign of the gravitational pressure $P$. 

Energy forms can also change into each other rapidly during a phase transition in the universe. Once such rapid phase transition occurred at the end of inflation, when dark energy, presumably $w=-1$ energy, rapidly changed into (mostly) $w = 1/3$ radiation energy. 

A leading term in the fast transition of $w=0$ matter to $w=1/3$ radiation is fusion in the central regions of stars. The power density is $d\rho_*/dt = n_* L_*$, where $n$ refers to average number density, $L$ refers to absolute luminosity, and the subscript * designates stars. Assume that there are $n_* \approx 10^{-69}$ galaxies \,m$^{-3} \times 10^{11}$ stars galaxy$^{-1}$, and that each star produces on average $L_* = 4 \times 10^{26}$\,kg\,m$^2$\,s$^{-3}$. The total average power converted would be on the order of $n_* L_* \approx 4 \times 10^{-32}$\,kg\,m$^2$\,s$^{-3}$. Given that the energy of a single hydrogen atom is about $m_P c^2 \approx 1.5 \times 10^{-10}$\,kg\,m$^2$\,s$^{-2}$, the average power transferred approximately corresponds to a conversion of one hydrogen atom to radiation over the volume of the Earth for every four seconds.

\section{The Past and Future According to the Friedmann Equations}

The Friedmann equations may be incorrect or incomplete. Other classes of solutions are occasionally considered. For example, a popular alternative to the classical Friedmann energy equation has $H^2 = A \rho + B \rho^b$, where $b<2/3$ is time dependent. A universe so described is labeled Cardassian.\citep{Fre03} For Friedmann cosmologies, $b = 0$. No exceptions to the Friedmann equations have been found on cosmological scales. So far the Friedmann equations have been able to explain the overall evolution of our universe. The history of the universe and its future, as extrapolated from the Friedmann equations, is discussed in the following section.

\subsection{A really short history of the universe in terms of dominating forms of energy}

The flatness of the universe and the homogeneity of the microwave background do not give us details about the past history of the universe. If these attributes are caused by the expansion history of the universe, they tell us that the universe must have spent a significant amount of time in a $w < -1/3$ mode to create the flatness and the homogeneity seen today. 

In this context a brief summary of the history of the universe is possible that highlights which stable energy forms dominated the Friedmann equations. A summary of this section is given in Table~\ref{tab3}. From current cosmological data the farthest back that the universe can be traced is to Planck time, an epoch where all the known forces had the same strength. The Planck time occurred at about $t_P \approx 10^{-43}$\,s after the Big Bang, a time where the density would be formally infinite. At the Planck time and for all times later, the universe expanded.\citep{Wik07}  Just after the Planck time, the universe was $w=1/3$ radiation dominated.

The general lack for the present need of any ($1 - \Omega_{\rm total}$) curvature term in the Friedmann equations indicates that the universe entered into at least one inflationary epoch where some sort of dark energy with $w<-1/3$ dominated in the past. When any energy form that has $w<-1/3$ dominates, $\Omega_{\rm total}$ for the universe moves toward unity. It is commonly assumed, though, and will be assumed here, that this dark energy had $w=-1$. If inflation was related to the grand unification energy scale of forces, it likely started at a time near $\approx 10^{-35}$\,s after the Big Bang.\citep{Pea99}

At the end of inflation, $w=-1$ energy decayed into a mixture of $w=-1$, $w=0$, and $w=1/3$ energy form, with the $w=1/3$ relativistic energy form dominating the energy density and determining the expansion rate.  This new ``radiation epoch" started at a time of about $10^{-32}$\,s after the Big Bang. 

As the universe further expanded, $w=1/3$ energy density diluted relative to the $w=0$ matter energy density and the still diminutive $w=-1$ energy density. In time, the $w=1/3$ radiation density diluted below the $w=0$ ($n=3$) matter energy density.  This ``matter epoch" started at a time of about 70,000 years after the Big Bang .

As the universe continued to expand, the $w=1/3$ radiation energy density further diluted to cosmological insignificance, as it remains today. Today, the $w=0$ matter energy density has now diluted so much that the remaining constant $w=-1$ dark energy density is again beginning to dominate the total energy density of the universe, and hence the expansion rate of the universe. 

\subsection{Big Rip, Big Freeze, or Big Crunch: Possible end states of the universe} 

If the universe remains dominated by stable forms of energy, there are only three possible end states. All of them are extrapolated from the Friedmann equations. The first is called the Big Rip where a pervasive and gravitationally repulsive phantom energy grows to infinite magnitude and rips everything apart. Another possibility is the Big Freeze, where the universe expands forever, everything slows down and cools, and eventually the universe becomes a sea of expanding static, stable energy units. The last possibility is the Big Crunch, where the universe expands for only a finite time and then re-collapses in a heat bath that compresses everything together toward infinite density. Which of these possible fates awaits our universe is discussed in the following sections and summarized in Table~\ref{tab4}.

If the dominant energy form somehow changes into other stable forms of energy, other futures for our universe are possible. These include oscillating universes, where the universe re-expands again after a Big Crunch,\citep{Ste05} and a cascading universe where a small phantom energy will grow to dominate the universe until it decays into daughter products that contain yet another phantom energy.\citep{Gru04}

\subsection{The future of a flat, minimal universe}

An instructive, minimal case occurs for a flat universe, when $\Omega = 1$ in only one form of energy. Here the Friedmann equations of energy and acceleration take on very simple forms, namely those shown in Eq.~(\ref{FEminall}) and their form-specific solutions are given in Section III.  The end states are summarized in Table~\ref{tab4}.

Possibly universe futures will be discussed starting at the lowest value of $w$ when a $w<-1$ phantom energy form dominates, the universe expands, the speed of expansion continually increases, and the acceleration of the expansion increases faster than exponentially. This universe will end in a finite time in a Big Rip.

When a $w=-1$ cosmological constant energy form dominates the universe, the speed of the expansion and the acceleration of the expansion continually increase exponentially. This universe will end in an infinite time in a Big Freeze. When a $w = -2/3$ domain wall energy form dominates, the speed of expansion continually increases, but the acceleration of the universe remains a positive constant. This universe will also end in an infinite time in a Big Freeze.

When a $w=-1/3$ cosmic string energy form dominates, the universe ``coasts" at constant expansion speed but zero acceleration. This universe will also end in an infinite time in a Big Freeze. When a $w=0$ particle energy form dominates the universe, the magnitude of the expansion speed continually decreases to zero, as does the magnitude of the acceleration of the expansion. The universe will asymptotically approach a Big Freeze. 
The same analyses holds for a $w=1/3$ radiation dominated universe, as well as an ultralight $w>1/3$, universe.

\subsection{The future of a curved, minimal universe}

We next consider universes dominated by a single energy form but with $\Omega \ne 1$ so that a curvature term is present. According to the Friedmann equation of energy, the future of the expansion speed of the universe might then become a competition between the term in Eq.~(\ref{generalequation}) that describes the evolution of this energy form and the trailing curvature term. Curvature will not affect the magnitude of the acceleration of the expansion. Table~IV summarizes these possibilities. 

Let's first look at energy forms that dilute slower than $a^{-2}$. These have $w<-1/3$ and will dominate the curvature term, which goes as $a^{-2}$. In these universes it does not matter what $\Omega_{\rm total}$ and the curvature is at any time -- the universe will always expand and qualitatively act as the $\Omega=1$ universe described earlier, ending either in a Big Rip or a Big Freeze.

For a universe dominated by cosmic strings, there is a tie between the energy term and the curvature term -- both go as $a^{-2}$. Qualitatively this universe will evolve as a $\Omega=1$ universe with no curvature, ending in a Big Freeze. For a universe dominated by matter, radiation, or ultralight, a finite curvature term will eventually grow to dominate any of these terms as the universe expands. If $(1 - \Omega_{\rm total}) > 0$, then $\Omega_{\rm total} < 1$, and the universe will expand forever and end in a Big Freeze. This case is called an ``open universe." 

If $(1 - \Omega_{\rm total})<0$, then $\Omega_{\rm total} > 1$, and eventually the expansion rate ${\dot a}$ will go to zero. At that time a maximum expansion scale factor of the universe can be found. Past this time, these universes will collapse into a Big Crunch.

It is also useful to understand that any universe that spends a significant amount of time being dominated by a $w < -1/3$ energy phase drives $\Omega_{\rm total}$ to unity. Alternatively, a universe that spends a significant time being dominated by a $w> -1/3$ phase drives $\Omega_{\rm total}$ away from unity. 

If present observations continue to support the concordance cosmology where roughly 70\% of the energy in the universe is in a form where $w=-1$, and if the energy forms in the universe today are stable, then the universe should continue to expand and the average $w$ of the universe will continue to approach $-1$. The fate of the universe will then be a Big Freeze. It is possible that the $w=-1$ energy will evolve as a field and$/$or decay into energy forms with higher $w$. If so, the ultimate future of the universe is unknown.

\begin{acknowledgements}
We thank Noah Brosch, Amir Shahmoradi, and two anonymous referees for helpful comments. 
\end{acknowledgements}

\section*{Tables}

\begin{table}[h]
\begin{center}
\begin{tabular}{| c | c | c | c |}
\hline
$w$ & $ n$ & Dimensional Type & Name \\
\hline 
$<-1$ & $<$0 & unknown & Phantom energy \\
\hline
$-1$ & 0 & volume & Cosmological constant \\
\hline
$-2/3$ & 1 & sheet & Domain wall \\
\hline
$-1/3$ & 2 & line & Cosmic string \\
\hline
0 & 3 & point & Matter \\
\hline
1/3 & 4 & relativistic point & Radiation \\
\hline
$>1/3$ & $>4$ & unknown & Ultralight \\
\hline
\end{tabular}
\caption{\label{tab1}Energy forms that may drive the expansion rate of the universe as described by the Friedmann equations. The entries include known and hypothesized energy forms that are thought to be stable perfect fluids over cosmological scales.}
\end{center}
\end{table}

\begin{table}[h]
\begin{center}
\begin{tabular}{| l | c | c | c | c | c | c | c | c |}
\hline
\vspace{-0.13in}
Energy & Phantom & Cosmic & Domain & Cosmic & Particle & Radiation & Ultralight \\
Form & Energy & Constant & Walls & Strings & Matter & Energy & Energy \\
\hline 
Phantom & Repel & Repel & Repel & Repel & Repel & Repel & -- \\
\hline
Cosmic constant & Repel & Repel & Repel & Repel & Repel & Repel & -- \\
\hline
Domain Walls & Repel & Repel & Repel & Repel & Repel & Neutral & Attract \\
\hline
Cosmic String & Repel & Repel & Repel & Neutral & Neutral & Attract & Attract \\
\hline
Matter & Repel & Repel & Repel & Neutral & Attract & Attract & Attract \\
\hline
Radiation & Repel & Repel & Neutral & Attract & Attract & Attract & Attract \\
\hline
Ultralight & -- & -- & Attract & Attract & Attract & Attract & Attract \\
\hline
\end{tabular}
\caption{\label{tab2}How do two stable energy forms act toward each other? The behavior is extrapolated from Ref.~\onlinecite{Vil81a}.}
\end{center}
\end{table}

\begin{table}[h]
\begin{center}
\begin{tabular}{| l | c | c | c |}
\hline
Name & Time (Approximate) & Dominant Energy & $w$s \\
\hline 
Planck Time & 10$^{-43}$\,s & Radiation & 1/3 \\
\hline
Inflation & $10^{-35}$\,s & Cosmic constant & $-1$ \\
\hline
Radiation Epoch & $10^{-32}$\,s & Radiation & 1/3 \\
\hline
Matter Epoch & 70,000 years & Matter & 0 \\
\hline
Modern Epoch & 13.7 billion years & Cosmic constant & $-1$ \\
\hline
\end{tabular}
\caption{\label{tab3}A brief history of the universe in terms of the stable energy forms that dominated the gravitational expansion in the Friedmann equations.}
\end{center}
\end{table}

\begin{table}[h]
\begin{center}
\begin{tabular}{| l | c | c | c |}
\hline
Dominating Energy Form & Equation of State & Universe Density & Universe Fate \\
\hline 
Phantom energy & $w < -1$ & $\Omega_{\rm total} = \mbox{any}$ & Big Rip \\
\hline
Cosmic constant, domain walls, or cosmic strings & 
$-1 \le w \le -1/3$ &
$\Omega_{\rm total} = \mbox{any}$ &
Big Freeze \\
\hline
Matter, radiation, or ultralight &
$w > -1/3$ &
$\Omega_{\rm total} \le 1$ &
Big Freeze \\
\hline
Matter, radiation, or ultralight &
$w > -1/3$ &
$\Omega_{\rm total} > 1$ &
Big Crunch \\
\hline
\end{tabular}
\caption{\label{tab4}The future of a universe dominated by a single energy form. Flat and curved universes are considered.}
\end{center}
\end{table}

\end{document}